 \documentclass[twocol]{ametsoc}

\journal{jpo}

\bibpunct{(}{)}{;}{a}{}{,}

\usepackage{comment}

\usepackage{doi}
\usepackage[dvipsnames]{xcolor}
\usepackage{hyperref}
\hypersetup{
	breaklinks=true,
	colorlinks=true,
	linkcolor=blue,
	citecolor=Purple,
	urlcolor=purple,
	pdfstartview={FitH},
	pdfview={FitH 0},
	pdfauthor={N C Constantinou},
	pdftitle={A barotropic model for eddy saturation},
 }

\usepackage{microtype}

\usepackage{xspace}

\newcommand{\MATLAB}{\textsc{Matlab}\xspace}

\newcommand{\beq}{\begin{equation} }

\newcommand{\eeq}{\end{equation} }

\newcommand{\defn}{\ensuremath{\stackrel{{\scriptscriptstyle\mathrm{def}}}{=}}}

\def\beq{\begin{equation}}
\def\eeq{\end{equation}}

\newcommand{\J}{\boldsymbol{\mathsf{J}}}

\newcommand{\bE}{\ensuremath {\boldsymbol {E}}}

\newcommand{\leta}{\ell_\eta}

\newcommand{\etarms}{\eta_{\mathrm{rms}}}

\providecommand\bnabla{\boldsymbol{\nabla}}
\providecommand\bcdot{\boldsymbol{\cdot}}

\def\wind{F}

\def\ii{{\rm i}}
\def\dd{{\rm d}}
\def\ee{{\rm e}}

\newcommand{\Mop}{\textsf{L}}
\newcommand{\Aop}{\textsf{A}}

\newcommand{\D}{\mathrm{D}}

\def\la{\langle}
\def\ra{\rangle}

\def\Fs{F/(\mu\etarms\leta)}
\def\Fs{F/(\leta\etarms^2)}

\newcommand{\lap}{\ensuremath{\triangle}}
\renewcommand{\lap}{\nabla^2}
\newcommand{\p}{\ensuremath{\partial}}
\newcommand{\half }{\tfrac{1}{2}}

\newcommand{\grad}{\ensuremath{\boldsymbol {\nabla}}}

\renewcommand{\J}{\mathsf{J}}
\renewcommand{\J}{\mathrm{J}} 
\renewcommand{\J}{\mathsf{J}}
\renewcommand{\D}{\mathsf{D}}
\newcommand{\com}{\, ,}
\newcommand{\per}{\, .}

\newcommand{\z}{\zeta}
\newcommand{\h}{\eta}

\newcommand{\hop}{\eta^{({\rm op.})}}
\newcommand{\hcl}{\eta^{({\rm cl.})}}

\renewcommand{\hop}{\eta^{\parallel}}
\renewcommand{\hcl}{\eta^{\times}}
\renewcommand{\hop}{\eta^{\parallel}}
\renewcommand{\hcl}{\eta^{\times}}

\newcommand{\lop}{\leta^{({\rm op.})}}
\newcommand{\lcl}{\leta^{({\rm cl.})}}
\renewcommand{\lop}{\leta^{\parallel}}
\renewcommand{\lcl}{\leta^{\times}}

\renewcommand{\(}{\left(}
\renewcommand{\[}{\left[}
\renewcommand{\)}{\right)}
\renewcommand{\]}{\right]}
\newcommand{\<}{\left\langle}
\renewcommand{\>}{\right\rangle}

\newcommand{\transp}{\textrm{T}}

\newcommand{\corr}{\mathop{\mathrm{corr}}}

\def\XXint#1#2#3{{\setbox0=\hbox{$#1{#2#3}{\int}$}
     \vcenter{\hbox{$#2#3$}}\kern-.5\wd0}}

\def\fs{\la\psi\p_x\eta\ra}
\def\fsb{\la\bar{\psi}\p_x\eta\ra}

\title{A barotropic model of eddy saturation}

\authors{Navid C. Constantinou\correspondingauthor{Navid Constantinou, Scripps Institution of Oceanography, University of California San Diego, La Jolla, CA 92093-0213, USA.}}

\affiliation{Scripps Institution of Oceanography, University of California San Diego, La Jolla, California, USA}

\ifdraft
  \email{navid@ucsd.edu}
\else
  \email{\href{mailto:navid@ucsd.edu}{navid@ucsd.edu}}
\fi

\abstract{``Eddy saturation'' refers to a regime in which the total volume transport of an oceanic current is insensitive to the wind stress strength. Baroclinicity is currently believed to be key to the development of an eddy-saturated state. In this paper, it is shown that eddy saturation can also occur in a purely barotropic flow over topography, without baroclinicity. Thus, eddy saturation is a fundamental property of barotropic dynamics above topography. It is demonstrated that the main factor controlling the appearance or not of eddy-saturated states in the barotropic setting is the structure of geostrophic contours, that is the contours of~$f/H$ of the ratio of the Coriolis parameter to the ocean's depth. Eddy-saturated states occur when the geostrophic contours are \emph{open}, that is when the geostrophic contours span the whole zonal extent of the domain. This minimal requirement for eddy-saturated states is demonstrated using numerical integrations of a single-layer quasi-geostrophic flow over two different topographies characterized by either open or closed geostrophic contours with parameter values loosely inspired by the Southern Ocean. {\color{black}In this setting, transient eddies are produced through a barotropic--topographic instability that occurs due tot the interaction of the large-scale zonal flow with the topography. Through the study of this barotropic--topographic instability insight is gained on how eddy-saturated states are established.}}

\begin{document}

\maketitle

\section{Introduction}

The Southern Ocean, and in particular the Antarctic Circumpolar Current (ACC), are key elements of the climate system. The ACC is driven by a combination of strong westerly winds and buoyancy forcing. \citet{Straub-1993} advanced the remarkable hypothesis that the equilibrated ACC zonal transport should be insensitive to the strength of the wind stress forcing. This insensitivity was later verified in eddy-resolving ocean models of the Southern Ocean and is now referred to as \emph{eddy saturation} \citep{Hallberg-Gnanadesikan-2001,Tansley-Marshall-2001,Hallberg-Gnanadesikan-2006,Hogg-etal-2008,Nadeau-Straub-2009,Farneti-etal-2010,Hogg-2010,Meredith-etal-2012,Nadeau-Straub-2012,Morisson-Hogg-2013,Munday-etal-2013,Nadeau-etal-2013,Abernathey-Cessi-2014,Farneti-etal-2015,Nadeau-Ferrari-2015,Howard-etal-2015,Marshall-etal-2017}. Some indications of eddy saturation are seen in observations \citep{Boning-etal-2008,Firing-etal-2011}.


There is evidence that the strength of the westerly winds over the Southern Ocean, which force the ACC, are increasing \citep{Thompson-Solomon-2002,Marshall-2003,Yang-etal-2007,Swart-Fyfe-2012}. Recently, \citet{Hogg-etal-2015} using satellite altimetry data identified that along with the strengthening of the westerlies comes also a linear trend of the Southern Ocean surface eddy kinetic energy (EKE) while the ACC zonal transport remains insensitive or even has decreased. This way, \citet{Hogg-etal-2015} concluded that the ACC is in an eddy-saturated state. Given the strengthening of the Southern Ocean westerly winds over the last decades, and the potential enhanced strengthening under global warming forcing \citep{Bracegirdle-etal-2013}, the question that naturally arises is how will the ACC transport respond? Thus, understanding the mechanisms behind eddy saturation is particularly relevant. This paper attempts to shed some more insight in the mechanism underlying eddy saturation.

{\color{black} Initially the explanation for eddy saturation given by \citet{Straub-1993} relied on baroclinic processes and on the existence channel walls. In the following detailed models of \citet{Nadeau-Straub-2009,Nadeau-Straub-2012}, \citet{Nadeau-etal-2013}, and \citet{Nadeau-Ferrari-2015} the arguments for explaining eddy saturation were barotropic in heart. Specifically, \citet{Nadeau-Ferrari-2015} argued that the circulation can be decomposed to a circumpolar mode and a gyre mode (with the latter not contributing to the total transport). \citet{Nadeau-Ferrari-2015} showed that the wind stress curl spins barotropic gyres 
and, furthermore, that an increase of wind stress spins up the gyres while leaves the circumpolar mode intact; thus the transport remains insensitive. Still, though, all those explanations relied in baroclinic instability for producing transient eddies to transfer the momentum down from the surface of the ocean to the bottom. On the other hand, \citet{Marshall-etal-2017} and \citet{Mak-etal-2017} recently showed that eddy saturation can emerge as a result of an eddy flux parametrization that was introduced by \citet{Marshall-etal-2012}. In agreement with \citet{Straub-1993}, \citet{Marshall-etal-2017} also relate the production of EKE to the vertical shear of the zonal mean flow. Again, baroclinic instability is identified as the main source of EKE. In this paper, we show that eddy saturation can be observed without baroclinic eddies, without channel walls and without any wind stress curl, i.e. without any gyres. }

According to \citet{Johnson-Bryden-1989}  different density layers are coupled  via interfacial form stress that transfers momentum downwards from the sea-surface  to the bottom. But at the bottom it is topographic form stress that transfers momentum from the ocean to the solid earth \citep{Munk-Palmen-1951}. Note that only the standing eddies result in time-mean topographic form stress. Thus, it seems reasonable that topography should play a dominant role in understanding what sets up the total vertically-integrated transport.  {\color{black} There is a consensus that topography acts as the main sink in the zonal momentum balance (in agreement with \citet{Munk-Palmen-1951}). But, can the topography also have an active role in setting up the momentum balance, e.g.,~by shaping the standing eddy field and its associated form stress?} 
\citet{Abernathey-Cessi-2014} {\color{black}argued in favor of the latter: they} showed that isolated topographic features result in localized absolute baroclinic instability {\color{black}(i.e.~baroclinic eddy growth in situ above the topographic features)} and an associated almost-barotropic standing wave pattern. Transient eddies are suppressed away from the topography. Furthermore, relative to the flat-bottom case, the thermocline is shallower and isopycnal slopes are smaller. {\color{black}Thus, the usual arguments assuming that the isopycnal slopes are so steep as to be  marginal with respect to the flat-bottomed baroclinic instability cannot explain the ACC equilibration (and thus neither eddy saturation) in this model configuration with localized topography. In addition, \citet{Thompson-NaveiraGarabato-2014}, and more recently \citet{Youngs-etal-2017}, further emphasized the role of the  standing eddies (or standing meanders) in setting up the momentum balance and ACC transport.} In this paper, we emphasize the role of the standing eddies and their form stress in determining transport in an eddy-saturated regime within a barotropic setting. 

Lately, there has been increasing evidence arguing for the importance of the barotropic processes in determining the ACC transport. \citet{Ward-Hogg-2011} studied the ACC equilibration using a multi-layer primitive equation wind-driven model with an ACC-type configuration starting from rest. Following turn-on of the wind a strong barotropic current forms within several days that is able to transfer most of the imparted momentum to the bottom; only after several years does the momentum start being transferred vertically via interfacial form stress. The fast response is that a bottom pressure signal arises  a few days after turn-on, and the associated topographic form stress couples the ocean to the solid earth. Subsequently, for about ten years, and contrary to the statistical equilibrium scenario  described by \citet{Johnson-Bryden-1989}, interfacial from stress transfers momentum vertically from the bottom \textit{upwards}. {\color{black}At equilibrium both eastward momentum is transfered from surface downwards and also westward momentum from bottom upwards.} On the other hand, studies using in situ velocity measurements, satellite altimetry, and output from the Southern Ocean State Estimate (SOSE) also argue in favor of the importance of the bottom-velocity component of the ACC transport, which comes about from the barotropic component of the flow\footnote{\color{black}Traditionally, the baroclinic component of the flow is obtained through the thermal-wind relationship after assuming zero flow at the bottom. Thus, any transport due to the bottom flow is not thought as part of the ``baroclinic component'' of the ACC transport.} \citep{Rintoul-etal-2014,PenaMolino-etal-2014,Masich-etal-2015,Donohue-etal-2016}. For example, using measurement from the cDrake experiment \citet{Donohue-etal-2016} estimated that the bottom-velocity component of the ACC transport accounts for about 25\% of the total transport.

\citet{Constantinou-Young-2017} discussed the role of standing eddies using a simple barotropic model forced by an imposed steady wind stress and retarded by a combination of bottom drag and topographic form stress \citep{Hart-1979,Davey-1980a,Holloway-1987,Carnevale-Frederiksen-1987}. \citet{Constantinou-Young-2017} used a random monoscale topography and argued that a critical requirement for eddy saturation is that the ratio of the planetary potential vorticity (PV) to the topographic PV  is large enough so that the geostrophic contours (that is the contours of the planetary plus the topographic contribution to~PV) are ``open'' in the zonal direction. Here, we demonstrate that what matters for eddy saturation is not the actual value of the ratio of the planetary potential vorticity (PV) over the topographic PV, but rather the structure of the geostrophic contours.

{\color{black}The main goal of the present paper is to provide insight on how eddy saturation is established in this barotropic setting. We do that by comparing numerical results using two simple sinusoidal topographies.} We show that this barotropic model without baroclinic instability can exhibit impressive eddy saturation provided that the geostrophic contours are open (see section~\ref{sec:results}). We do not claim here that baroclinic processes are not important for setting up the momentum balance in the Southern Ocean. Instead, we emphasize the role of barotropic dynamics and the fact that we can still observe eddy saturation without baroclinicity. In this barotropic model, transient eddies arise as an instability due to the interaction of the large-scale zonal flow with the topography. {\color{black}The simple topography used here allows us to study in detail the barotropic--topographic instability that gives rise to transient eddies and thus provide insight on how eddy-saturated states appear (see section~\ref{sec:stability}). We show, in this way, that topography does not only have a passive role in setting the zonal momentum balance acting as the main sink of zonal momentum, but it can also have an active role by producing transient eddies that shape the standing eddy field and its associated form stress.}

\section{Setup \label{sec:form}}

Consider the quasi-geostrophic dynamics of a barotropic fluid of depth $H-h(x,y)$ on a beta-plane. The fluid velocity consisting of a large-scale domain-averaged zonal flow, $U(t)$, along the zonal $x$ direction plus smaller-scale eddies with velocity $(u,v)$. The eddy component of the flow is derived via an eddy streamfunction $\psi(x,y,t)$ through $(u,v) = (-\p_y\psi, \p_x\psi)$; the  streamfunction of the total flow is $-U(t) y + \psi(x,y,t)$. The relative vorticity of the flow is $\lap\psi$, where $\lap \defn \p_x^2+\p_y^2$, and the quasi-geostrophic potential vorticity (QGPV) of the flow is
\beq
	f_0+\beta y +\h + \lap\psi \per \label{beq11}
\eeq
In~\eqref{beq11}, $f_0$ is the Coriolis parameter at the center of the domain, $\beta$ is the meridional PV gradient and $\h(x,y)\defn f_0 h(x,y)/H$ is the topographic contribution to QGPV or simply the \emph{topographic PV}. The QGPV and the large-scale flow evolve through:\begin{subequations}\label{EOM}
\begin{gather}
	\p_t \lap\psi + \J(\psi - U y\,,\lap\psi+\eta+\beta y)=- \D \lap\psi \com\label{eq:z_NL}\\
 \p_t U =\wind -\mu U  - \fs\com\label{eq:U_NL}
\end{gather}\end{subequations}
\citep{Hart-1979,Davey-1980a,Holloway-1987,Carnevale-Frederiksen-1987}. In~\eqref{eq:z_NL}
$\J$~is the Jacobian, $\J(a,b)\defn (\p_x a) (\p_y b) - (\p_y a)(\p_x b)$, and $\D\defn\mu + \nu_4\nabla^4$ is the dissipation operator, which includes linear Ekman drag~$\mu$ and hyperviscosity~$\nu_4$ used for numerical stability. On the right of~\eqref{eq:U_NL}, $\fs$ is the topographic form stress, with angle brackets denoting an average over the $(x,y)$-domain. The large-scale flow $U(t)$ in~\eqref{eq:U_NL} is forced by the constant $\wind=\tau/(\varrho_0 H)$, where $\tau$ is the uniform surface zonal wind stress and $\varrho_0$ is the reference density of the fluid, while is being retarded by a combination of bottom drag and topographic form stress. {\color{black}Being constant through the domain, the wind stress has no curl.} The domain is periodic in both the zonal and the meridional directions, with size $2 \pi L \times 2 \pi L$. Following \citet{Bretherton-Karweit-1975}, we have in mind a mid-ocean region that is smaller than ocean basins but much larger than the length scale of ocean macroturbulence. The role of hyperviscosity is limited only to the removal of small-scale enstrophy. Thus the hyperviscosity has a very small effect on larger scales and energy dissipation is mainly due to drag~$\mu$.

The model formulated in~\eqref{EOM} is the simplest model that can be used to investigate beta-plane turbulence above topography driven by a large-scale zonal wind stress applied at the surface of the fluid. It has been used in the past for studying the interaction of zonal flows with topography \citep{Hart-1979,Davey-1980a,Holloway-1987,Carnevale-Frederiksen-1987} and recently by \citet{Constantinou-Young-2017} for studying the geostrophic flow regimes above random monoscale topography.

Inspired by the Southern Ocean we take $L=775\;\textrm{km}$, $H=4\,\textrm{km}$, $\varrho_0=1035\;\textrm{kg}\;\textrm{m}^{-3}$, $f_0=-1.26\times10^{-4}\;\textrm{s}^{-1}$ and  $\beta=1.14\times10^{-11}\;\textrm{m}^{-1}\textrm{s}^{-1}$. Also, we take $h_{\rm rms}\defn\sqrt{\la h^2\ra}=200\;\textrm{m}$, which implies that $\etarms=6.3\times10^{-6}\;\textrm{s}^{-1}$. For Ekman drag we use $\mu=6.3\times10^{-8}\;\textrm{s}^{-1}\approx (180\; \textrm{day})^{-1}$ \citep{Arbic-Flierl-2004}.

Results are reported using two sinusoidal topographies:\begin{subequations}\begin{align}
\hop &= \sqrt{2}\etarms\,\cos\(14\,x/L\)\com\label{def_hop}\\
\hcl &= 2\etarms\,\cos\(10\,x/L\)\cos\(10\,y/L\)\per
\end{align}\label{topos}\end{subequations}
Both topographies in~\eqref{topos} are monoscale, that is they are characterized by a single length-scale $\leta= \etarms\big/|\grad\h|_{\rm rms}$, which is:
\begin{gather*}
  \lop \approx\lcl = 0.071\,L = 55.03\;\textrm{km}\per
\end{gather*}
Thus, for both topographies the ratio of planetary PV to topographic PV is $\beta /|\grad\h|_{\rm rms}=0.1$.

\begin{figure}
 \centerline{\includegraphics[width=17pc]{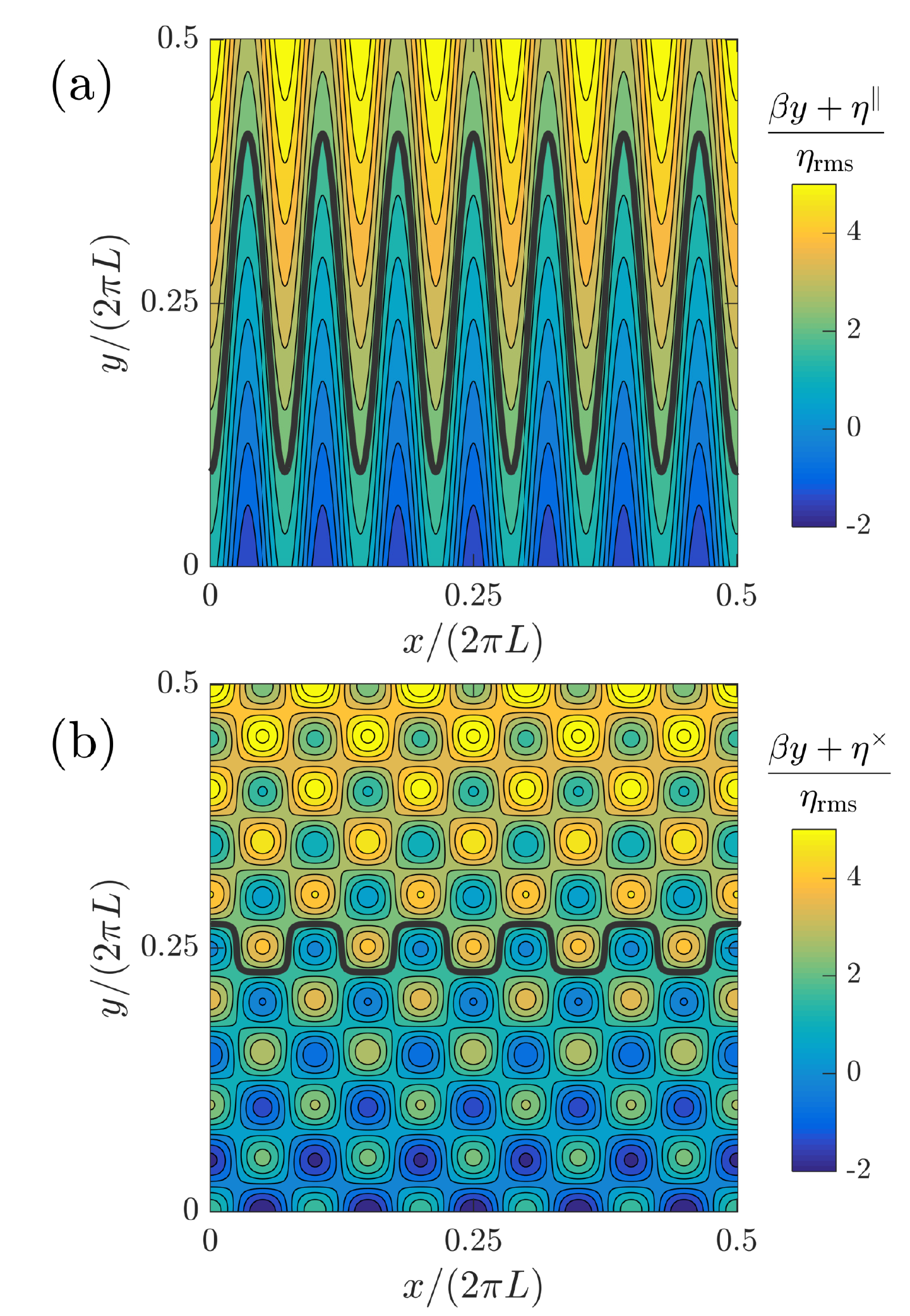}}
 \caption{The structure of the geostrophic contours, \protect$\beta y + \eta$, for the two types of topographies used in this paper: panel~(a) uses \protect$\hop$ and panel~(b) uses $\hcl$. For both cases $\beta\leta/\etarms=0.1$. An open geostrophic contour for each case is marked with a thick curve. The case $\beta y+\hop$ consists of only open geostrophic contours while $\beta y+\hcl$ consists of mostly closed ones. (Only a quarter of the flow domain is shown.)}\label{figtopo}
\end{figure}

An important factor controlling the behavior of the flow is the geostrophic contours, that is the level sets of $\beta y+\eta(x,y)$. Figure~\ref{figtopo} shows the structure of geostrophic contours for the two topographies $\hop$ and $\hcl$ using the parameters given above. Here, we distinguish between ``open'' and ``closed'' geostrophic contours: open geostrophic contours span the full zonal extent of the domain; see, for example, the two contours marked with thick black curves in each panel of figure~\ref{figtopo}. For any nonzero $\beta$, all geostrophic contours for topography $\hop$ are open while for topography $\hcl$ there exist both closed as well as open geostrophic contours. However, for topography $\hcl$ open geostrophic contours, such as the one shown in figure~\ref{figtopo}(b), are only found in the vicinity of narrow channels that span the horizontal extent of the domain snaking around local maxima and minima of $\beta y+\hcl$. We say, therefore, that $\hcl$ is characterized by closed geostrophic contours.

It is useful to decompose the eddy streamfunction $\psi$ into time-mean ``standing eddies'' with streamfunction $\bar \psi$,  and  residual ``transient eddies'' $\psi'$, so that:
\beq
	\psi(x,y,t) = \bar \psi(x,y) + \psi'(x,y,t)\com \nonumber\label{tStand}
\eeq
where the time-mean is $\bar \psi \defn \lim_{T\to\infty}\frac{1}{T} \int_{t_0}^{t_0+T}\!\!\! \psi(x,y,t') \, \dd t'$, with $t_0$ the time the flow needs to reach a statistically steady state. Similarly, all fields can be decomposed into time-mean and transient components e.g., $U(t) = \bar U + U'(t)$. We are interested on how the time-mean large-scale flow~$\bar U$, which is directly related to the time-mean total zonal transport, depends on wind stress forcing~$F$.

All solutions presented in this paper employ $512^2$ grid points; \textcolor{black}{this resolution allows about~6 grid points within~$\leta$}. We have verified that results remain unchanged with double the resolution. The hyperviscocity coefficient is set to $\nu_4=2.27\times10^{9}\;\textrm{m}^{4}\,\textrm{s}^{-1}$. System~\eqref{EOM} is evolved using the Exponential Time Differencing 4th-order Runge-Kutta time-stepping scheme~\citep{Cox-Matthews-2002,Kassam-Trefethen-2005}.\footnote{A \MATLAB code used for solving~\eqref{EOM} is available at the github repository: \url{https://github.com/navidcy/QG_ACC_1layer}.} Time-averaged quantities are calculated by averaging the fields over the interval $30\le\mu t\le 60$.

\begin{figure}
 \centerline{\includegraphics[width=19pc]{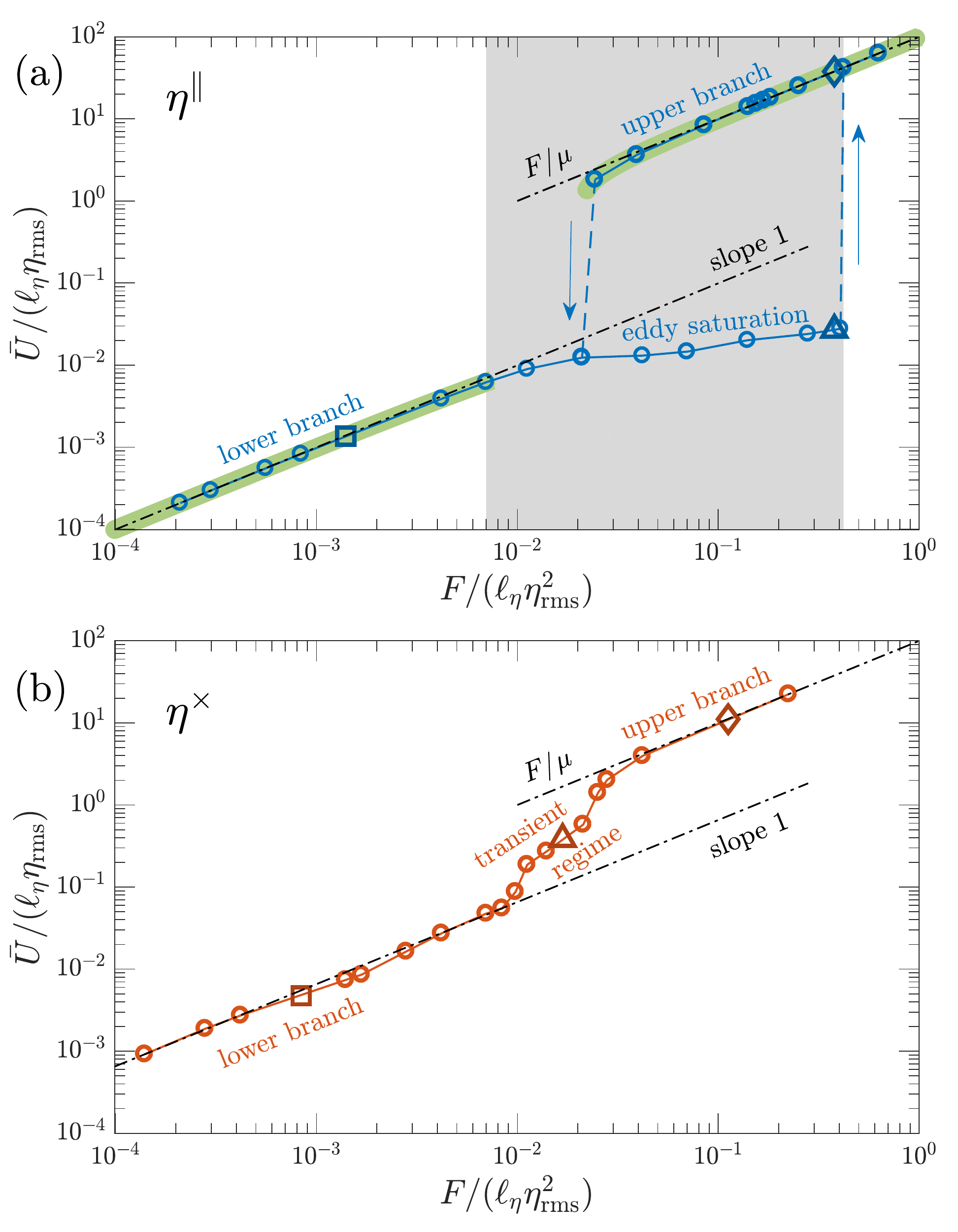}}
 \caption{The equilibrated time-mean large-scale flow $\bar U$ as a function of the non-dimensional forcing for the two topographies. Panel~(a) shows the case with open geostrophic contours while panel~(b) shows the case with closed geostrophic contours. Dash-dotted lines mark the slope~1 and the time-mean large-scale flow $\bar U=\wind/\mu$. In panel~(a) the eddy saturation regime $7.0\times10^{-3}\le \Fs \le 4.1\times10^{-1}$ is shaded. The thick semitransparent green curve marks the stability region of the steady solution~\eqref{eq:UeCeSe}. The points marked with a square~$\square$, a triangle~$\vartriangle$ and a diamond~$\lozenge$ correspond to the three typical cases for which the energy spectra and flow field snapshots are shown in figures~\ref{spec_open} and~\ref{spec_closed}.}\label{U_F_comb}
\end{figure}

\section{Results\label{sec:results}}

\subsection{Variation of the time-mean large-scale flow with wind stress forcing}

Figure~\ref{U_F_comb} shows how the time-mean large-scale flow $\bar U$ varies with wind stress forcing $F$ for the two topographies. A generic feature of model~\eqref{EOM}, for both topographies used here (as well as other more complex and/or multiscale topographies) is the following: as wind stress strength varies from weak to stronger values the flow transitions from a regime with relatively small time-mean large-scale flow $\bar{U}$ to a regime with large $\bar{U}$. Specifically, for very high wind stress forcing values the time-mean large-scale flow becomes $\bar{U}\approx F/\mu$, which is the value for $\bar{U}$ in the flat-bottom case (cf.~\eqref{eq:U_NL} with $\h=0$). The flat-bottom transport value is marked with the dashed-dotted line in both panels of figure~\ref{U_F_comb}. These two flow regimes are referred to as the lower-branch and the upper-branch \citep{Davey-1980a}, and they are indicated in figure~\ref{U_F_comb}. In both branches, the flow reaches a steady state without any transients after an initial adjustment on the time scale of $O(\mu^{-1})$. For intermediate wind stress forcing values the flow develops transients and becomes turbulent.

In the remainder of this subsection we describe in detail the qualitative features of the flow on the lower and upper branches, as well as the transition from one branch to the other for the two topographies.

\vspace{1em}

\begin{figure}
 \centerline{\includegraphics[width=19pc]{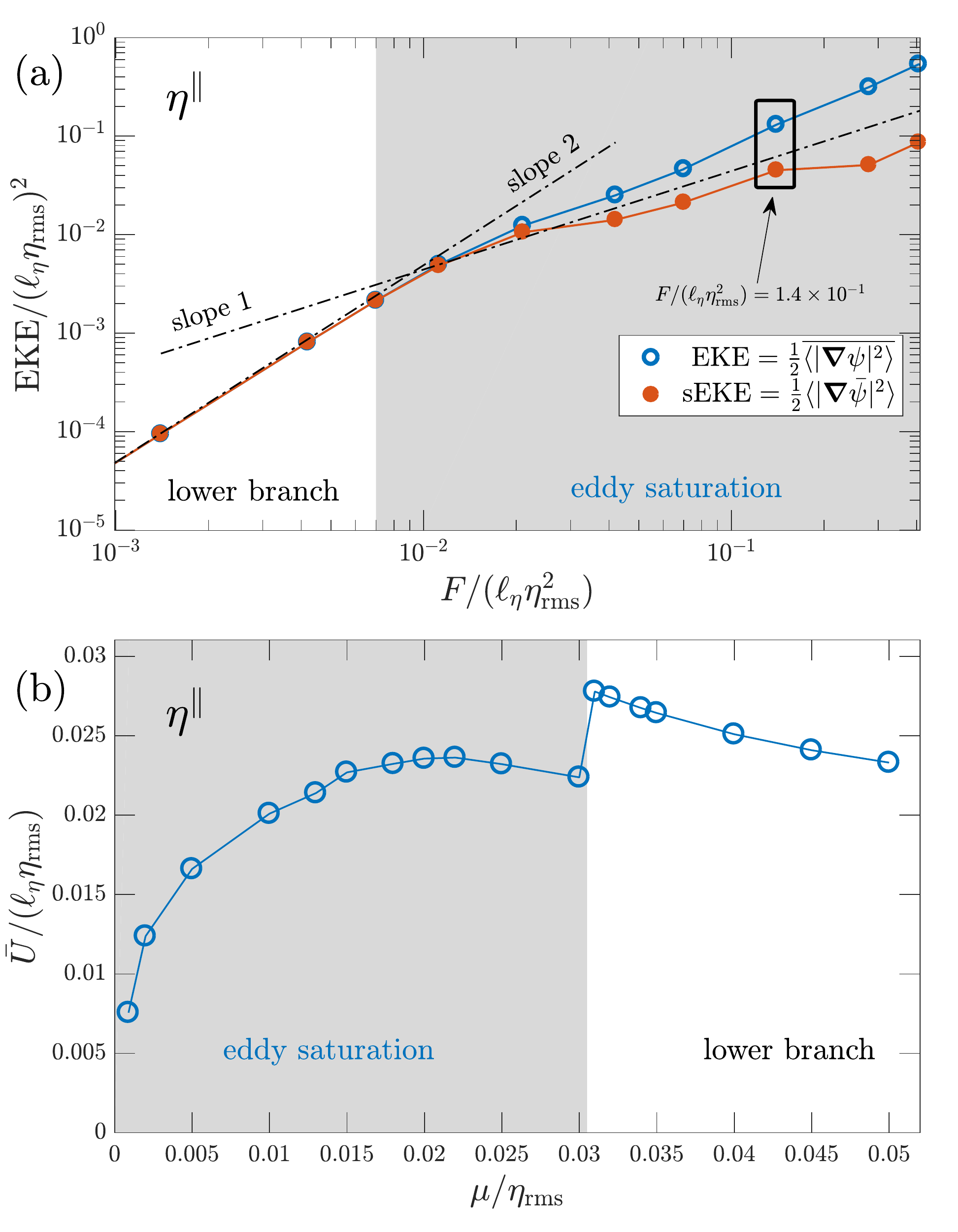}}
 \caption{Panel~(a) shows how the equilibrated EKE and the standing wave EKE (sEKE) vary with wind stress forcing for the topography $\hop$, as the flow transitions from the lower-branch to the eddy saturation regime. Dash-dotted lines mark the slopes~1~and~2. Panel (b) shows how the equilibrated time-mean large-scale flow~$\bar U$ varies with bottom drag~$\mu$ in the eddy saturation regime. For these experiments wind stress forcing is kept constant at $\Fs=1.4\times10^{-1}$ (highlighted in panel~(a)) while bottom drag $\mu$ varies.}\label{fig:EKEmu}
\end{figure}

For weak wind stress forcing values (i.e.~on the lower branch) the equilibrated solutions of~\eqref{EOM} are time-independent without any transient eddies. These steady states have a large-scale flow $U$ and an associated stationary eddy flow field $\psi$ that both vary linearly with wind stress forcing~$F$. As a result, EKE varies quadratically with wind stress, i.e., as~$F^2$. Figure~\ref{fig:EKEmu}(a) shows how EKE varies with wind stress forcing for topography $\hop$ and confirms the $F^2$ dependence on the lower branch (similar behavior is found for topography $\hcl$; not shown). As the wind stress increases beyond a certain value the lower-branch steady states undergo an instability and develop transients. Even though the topography used here imposes a length-scale $\ell_\eta$, the transient eddies can have scales much smaller than $\leta$. The onset of this instability is roughly at $\Fs\approx 7\times10^{-3}$ for both topographies. (The stability of the lower-branch states for topography $\hop$ is studied in detail in section~\ref{sec:stability}.)

For wind stress forcing values beyond the threshold of the lower-branch instability the flow above the two topographies are qualitatively very different: topography $\hop$ shows eddy saturation while topography $\hcl$ does not (see figure~\ref{U_F_comb}).

For topography $\hop$, as wind stress continues to increase beyond the onset of the transient eddy instability, the time-mean large-scale flow $\bar U$ ceases to grow linearly with $F$; instead $\bar U$ grows at a much slower rate. In fact, $\bar U$ shows only a very weak dependence on wind stress strength in the shaded region of figure~\ref{U_F_comb}(a): $\bar U$ increases only about \mbox{4-fold} in the course of a \mbox{60-fold} wind stress increase from $\Fs=7.0\times10^{-3}$ up to~$4.1\times1  0^{-1}$. This regime is identified as the \emph{eddy saturation regime}.\footnote{{\color{black}Note, that \citet{Nadeau-Ferrari-2015} found a similar weak dependence of the transport to wind stress strength in a series of experiments they performed without wind stress curl, i.e. similar to the forcing we used here. In their experiments with wind stress curl the eddy saturation was more pronounced.}}

The eddy saturation regime in figure~\ref{U_F_comb}(a) terminates at $\Fs\approx4.1\times10^{-1}$ by a discontinuous transition to the upper-branch, marked with a dashed vertical line. The upper branch steady solutions are characterized by much larger values of time-mean large-scale flow $\bar U$ compared to ones in the eddy saturation regime or on the lower-branch. This discontinuous transition has been coined {\it drag crisis} \citep{Constantinou-Young-2017}. 
The location of the drag crisis exhibits hysteresis \citep{Charney-DeVore-79}. As we increase wind stress forcing the drag crisis and the transition to the upper branch occurs at $\Fs\approx4.1\times10^{-1}$. However, if we initiate our model with the upper-branch solutions and start decreasing the wind stress forcing, we can find solutions on the upper branch down to $\Fs\approx2.44\times10^{-2}$:~see figure~\ref{U_F_comb}(a). We note that to obtain the upper-branch solutions we have to be very delicate in initiating the system appropriately. Thus, although the upper-branch solutions shown in figure~\ref{U_F_comb}(a) are linearly stable, their basin of attraction is smaller than those of the co-existing eddy-saturated solutions. The extent of the region of multiple stable equilibria is explained by studying the stability of the upper-branch solution (see section~\ref{sec:stability}).

\ifdraft\else\vspace{1em}\fi

On the other hand, the case with topography with closed geostrophic contours $\hcl$ in figure~\ref{U_F_comb}(b), does not exhibit any drag crisis nor any multiple equilibria. Also, the simulations using topography~$\hcl$ do not show any sign of eddy saturation; there is only a slight, barely noticeable, decrease from the linear growth of the time-mean large-scale flow $\bar U$ in the range $8.0\times10^{-4}\le \Fs\le3.0\times10^{-3}$, and for those wind stress values the flow does not have any transient eddies. Transient eddies appear for wind stress forcing $\Fs>4.2\times10^{-3}$. For wind stress forcing values beyond the onset of transient eddies the flow transitions from the lower to the upper branch in a continuous manner between $\Fs\approx8.0\times10^{-3}$ and~$4.2\times10^{-2}$ and at a rate much faster than linear.

\begin{figure*}
\centerline{\includegraphics[width=39pc]{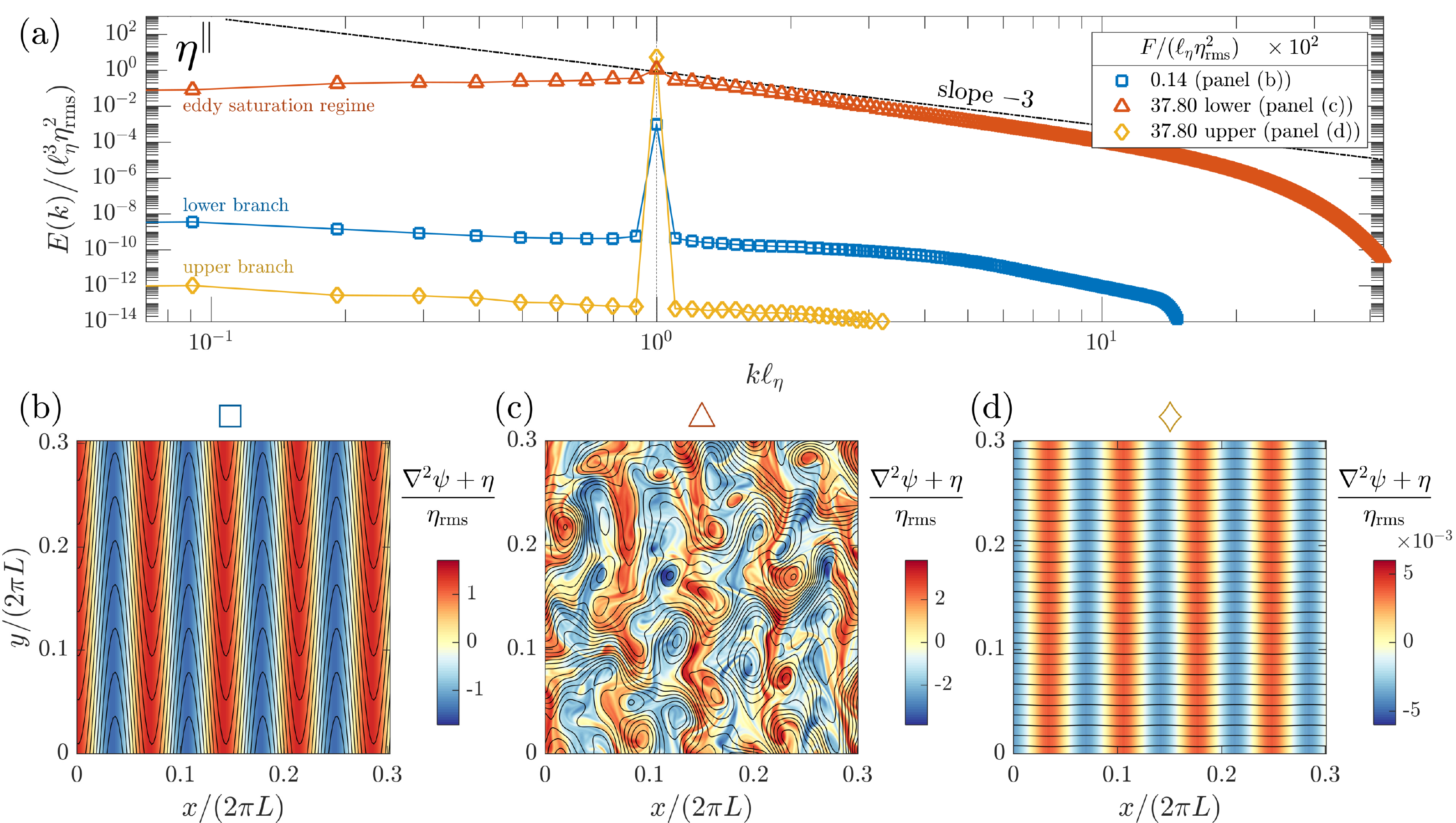}}
\caption{Panel (a) shows the energy spectra for three typical cases using topography $\hop$ with open geostrophic contours. The dash-dotted line marks the slope $-3$. Panels~(b),~(c),~and~(d) show a final snapshot of the sum of the relative vorticity and the topographic PV, $\lap\psi+\h$ (colors) overlayed with the total streamfunction, $\psi-Uy$ (contours), for each of the three cases presented in panel~(a). (In panels (b)-(d) only a fraction of the flow domain is shown for better visualization.)}\label{spec_open}
\ifdraft
	\end{figure*}
	\begin{figure*}
\else
	\vspace{1em}
\fi
\centerline{\includegraphics[width=39pc]{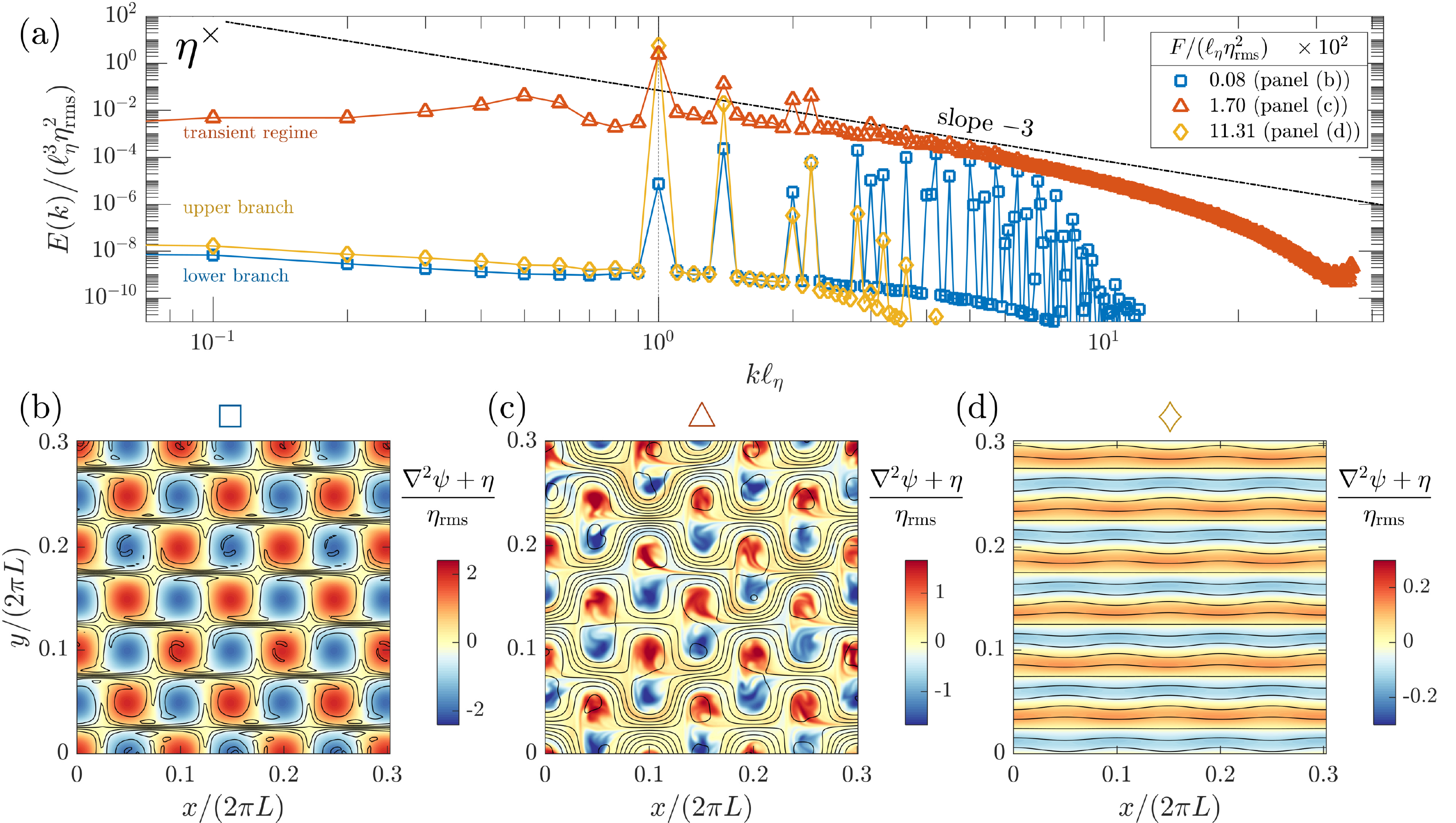}}
 \caption{Same as figure~\ref{spec_open} but for three typical cases using topography $\hcl$ with closed geostrophic contours.}\label{spec_closed}
\end{figure*}

\subsection{The flow regimes}

As described in the previous subsection, we distinguish three qualitatively different flow regimes for each topography. There exist, for both topographies, a lower-branch flow regime for weak wind stress forcing and an upper-branch flow regime for strong wind stress forcing. These flow regimes consist of steady flows without any transient eddies. In between the lower- and upper-branch flow regimes there exists a regime in which the flow has a transient component: the ``eddy saturation regime'' for topography $\hop$ case and the ``transient regime'' for topography~$\hcl$. These intermediate regimes are characterized by flows that feature strong transients and are turbulent, especially the eddy saturated case. Moreover, in these regimes the flow shows energy in a wide range of spatial scales. {\color{black}Figures~\ref{spec_open} and~\ref{spec_closed} show the energy spectra and a snapshot of the flow fields for a typical representative case of each of the three flow regimes (the cases presented in figures~\ref{spec_open} and~\ref{spec_closed} are the ones marked with a square, a triangle and a diamond in panels~(a) and~(b) of figure~\ref{U_F_comb}).}

The eddy field $\psi$ in the eddy saturated regime is characterized as two-dimensional turbulence. The wind stress~$F$ directly drives the large-scale flow $U$ in~\eqref{eq:U_NL} and, in turn, $U$ drives the eddy field through the term~$U\p_x\h$ in~\eqref{eq:z_NL}. This leads to a forward transfer of energy from the largest possible scale to the eddies on the length scale of the topography~$\leta$ {\color{black}and then nonlinearity transfers energy from length scale $\leta$ to all other scales}. This turbulent regime is anisotropic; for example, for the case marked with a triangle in figure~\ref{U_F_comb}(a) and also shown in figure~\ref{spec_open}(c), the transient eddy velocities are related by $\overline{\la u'^2\ra}\approx 0.57\overline{\la v'^2\ra}$. The {\color{black} turbulent eddy flow exhibits regimes that resemble the energy and enstrophy inertial ranges. For length scales smaller than $\leta$, the energy spectrum even shows a slope close to~$-3$ resembling the enstrophy inertial range as predicted by homogeneous isotropic turbulence arguments.} 
Characteristically, for the open geostrophic contours case (figure~\ref{spec_open}) the eddy saturated solution shows energy content in spatial scales other than $\leta$ that is at least eight orders of magnitude larger compared to the steady lower- and upper-branch solutions. Similar behavior is seen for the closed geostrophic contours case of figure~\ref{spec_closed}. The difference between these two cases is that for topography $\hcl$ the lower- and upper-branch solutions are not monochromatic, i.e.,~they do not show energy content only at scale~$\leta$: the nonlinearity due to non-vanishing Jacobian term $\J(\psi,\hcl)$ induces energy to cascade to scales smaller than scale $\leta$.

\subsection{Further characteristics of the eddy saturation regime for topography $\hop$}

{\color{black}It has been noted that in the eddy-saturation regime, even though the total ACC transport varies very little with wind stress, the domain-averaged EKE is approximately linearly related to the wind stress (e.g., see~\citet{Hallberg-Gnanadesikan-2001,Hallberg-Gnanadesikan-2006}). Figure~\ref{fig:EKEmu}(a) shows how this is reflected in this barotropic model: indeed,} in the eddy saturation regime the EKE varies with wind stress forcing at a rate much slower than quadratic. It is also apparent from figure~\ref{fig:EKEmu}(a) that strong transient eddies characterize the eddy saturation regime; this can be seen by the diminishing of the standing wave EKE (sEKE), that is the EKE that results from the standing waves alone.

The barotropic model~\eqref{EOM} predicts that for a fully eddy-saturated state, i.e.~when $\bar{U}$ does not vary at all with $F$, the EKE should vary linearly with wind stress forcing. {\color{black} If the flow is completely saturated, then from~\eqref{eq:U_NL} we get that the form stress $\fsb$ varies linearly with $F$. With that in hand, we can understand why EKE varies linearly with wind stress from the energy power integral, i.e.~$\overline{\la -\psi\times \text{\eqref{eq:z_NL}} \ra}$ gives:
\beq
\bar{U} \fsb +\underbrace{\overline{U'\la\psi'\p_x\h\ra}}_{\text{negligible}}=2\mu\underbrace{\la\half\overline{|\grad\psi|^2}\ra}_{\defn\,\text{EKE}} +\; \substack{\text{small hyperviscous}\\ \text{dissipation}}\;\;.\label{eq:energ}
\eeq
If we consider the energy transfers amongst the various flow components, we deduce that $\overline{U'\la\psi'\p_x\h\ra}$ should be less than $\bar{U} \fsb$ \citep{Constantinou-Young-2017}. Moreover, the numerical solutions of~\eqref{EOM} suggest that in the eddy saturation regime the ratio of $\overline{U'\la\psi'\p_x\h\ra}$ to $\bar{U} \fsb$ is less than about $10^{-3}$ and, therefore, negligible. Thus,~\eqref{eq:energ} implies that EKE is proportional to form stress $\fsb$, i.e.~EKE varies linearly with $F$.}



\vspace{1em}

{\color{black}Results both from quasi-geostrophic models \citep{Hogg-Blundell-2006,Nadeau-Straub-2012,Nadeau-Ferrari-2015} and also primitive-equation models \citep{Marshall-etal-2017} demonstrated the somehow counterintuitive fact that in the eddy saturation regime the total transport increases with increasing bottom drag. \citet{Hogg-Blundell-2006} suggested that this effect comes about because increasing the bottom drag decreases the strength of the transient eddies that are responsible for transferring momentum to the bottom where topographic form stress act; thus the strength of the zonal current increases. \citet{Nadeau-Straub-2012} on the other hand, argued that increase of the bottom drag damps the gyre circulation and its associated form stress; thus the zonal current increases to provide the drag needed to balance the momentum imparted by the wind stress.}

In the barotropic model studied here, we also show that transport increases with increasing bottom drag in the eddy saturation regime. Figure~\ref{fig:EKEmu}(b) shows how the time-mean large-scale flow $\bar U$ varies when wind stress forcing is kept fixed at $\Fs=1.4\times10^{-1}$ while bottom drag~$\mu$ varies: we find that $\bar U$ increases with $\mu$ (up to a point). {\color{black}Here the forcing is a steady mean zonal wind stress with no curl and thus large gyres do not form. Therefore, our results line-up more with the explanation given by \citep{Hogg-Blundell-2006} without, however, precluding the explanation by \citet{Nadeau-Straub-2012}.} Within model~\eqref{EOM} large bottom drag damps the eddy field $\psi$, thereby decreasing the form stress. Hence, increased bottom drag strengthens the time-mean large-scale flow $\bar U$, which is necessary to balance the wind stress forcing $F$ in the eddy saturation regime (cf.~\eqref{eq:U_NL}). If we increase bottom drag $\mu$ beyond a certain threshold the flow becomes laminar and transitions from the eddy-saturation regime to the lower-branch solution (for the case in figure~\ref{fig:EKEmu}(b) this transition occurs at $\mu/\etarms=3\times10^{-2}$). On the lower-branch the time-mean large-scale flow~$\bar U$ decreases with increasing~$\mu$.

\begin{figure*}
\centerline{\includegraphics[width=30pc]{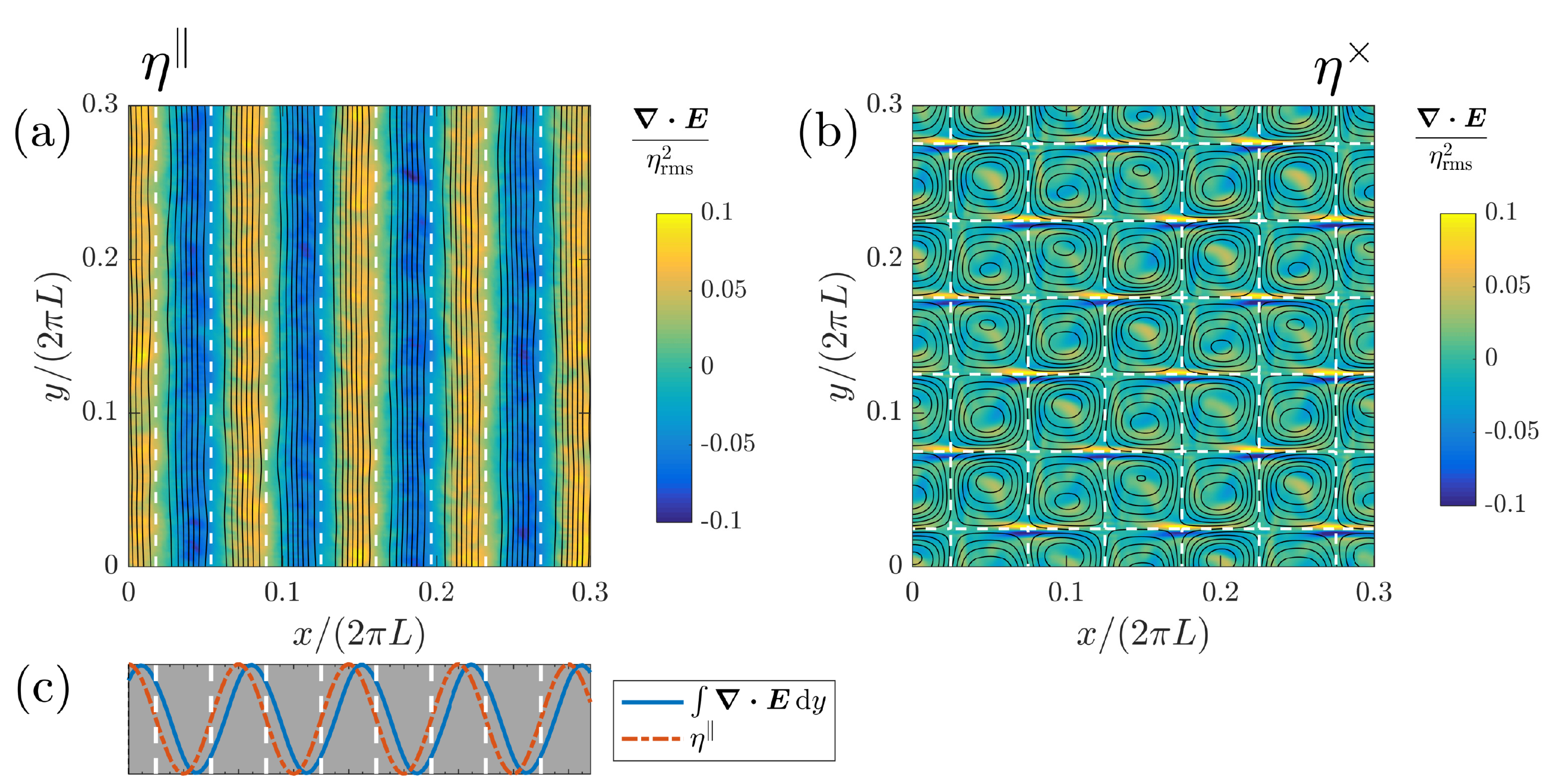}}
\caption{A comparison of the eddy flux divergence $\bnabla\bcdot\bE$ (shading) with the standing eddy streamlines $\bar{\psi}$ (black contours) for the two examples marked with a triangle~$\vartriangle$ in figure~\ref{U_F_comb} and also shown in figures~\ref{spec_open}(c) and~\ref{spec_closed}(c). For the open geostrophic contours case in panel~(a) the eddy PV fluxes dominate the whole domain and aligned with the mean flow streamlines. Panel~(c) compares the normalized $y$-integrated eddy PV flux divergence with the topography for the case shown in panel~(a).  Notice how the eddy PV flux divergence is offset by 1/8-th wavelength from the topography, thus exciting the $\sin{(mx)}$ component of the streamfunction. In all panels dashed white lines mark the $\h=0$ contour.}\label{fig:divE}
\end{figure*}

\section{The effect of transient eddies to the standing wave}

The eddy PV fluxes
\beq
\bE \defn \(\overline{(U'+u')\zeta'}\,,\,\overline{v'\z'}\) \com
\eeq
appears as forcing in the standing wave field~$\bar\psi$. To elucidate how the zonal momentum balance is shaped by the transient eddies for the two different topographies, we compute the divergence of the eddy PV fluxes, $\bnabla\bcdot\bE$. Figure~\ref{fig:divE} compares the structure of the eddy PV flux divergence (shading) with the actual standing wave~$\bar\psi$ (contours), for the two example cases shown in figures~\ref{spec_open}(c) and~\ref{spec_closed}(c).  There is an apparent qualitative difference between the two cases. For the open geostrophic contours case in panel~(a) the eddy PV fluxes are dominant over the whole domain and, to a large extent, they are aligned with the standing wave streamlines. This is not at all the case for the topography with closed geostrophic contours in panel~(b). In model~\eqref{EOM} we can verify, by considering the energy budgets among components \citep{Constantinou-Young-2017}, that the only source of transient eddy energy is the energy conversion from the standing-wave component through the term $\<\bar\psi\,\bnabla\bcdot\bE\>$. Thus, the stronger the transient eddies are the more stronger the correlation between eddy PV flux divergence and time-mean streamlines should be. In section~\ref{sec:stability} we show that for the case shown in figure~\ref{fig:divE}(a) there is a strong transient-eddy instability that explains this almost-perfect alignment between eddy PV fluxes and standing wave streamlines.

More importantly, however, notice how for the case with $\cos{(14x/L)}$-topography shown in figure~\ref{fig:divE}(a) the eddy PV fluxes are offset from the topography contours by 1/8-th wavelength (see figure~\ref{fig:divE}(c)). This offset in the eddy forcing induces a standing flow that projects onto the $\sin{(14x/L)}$ rather than the $\cos{(14x/L)}$, i.e., projects onto $\partial_x\hop$ rather than $\hop$. Since the time-mean topographic form stress is $\fsb$, it is apparent that only the $\sin$-component of the standing wave can produce non-zero topographic form stress. Thus, the mis-alignment between the topography and the transient-eddy forcing, induces a standing-eddy field which contributes to the form stress.

In conclusion, the transient eddies have the ability to shape the standing-wave field so that it projects more onto~$\partial_x\eta$ and, therefore, enhancing the time-mean topographic form stress. This shaping of the standing-wave field by the transient eddies is prominent in figure~\ref{fig:divE}(a) with topography $\hop$ but not in the case with $\hcl$. Specifically, for the case shown in figure~\ref{fig:divE}(a) the off-set by 1/8-th wavelength implies that $\bnabla\bcdot\bE$ projects equally\footnote{A mis-alignment of 1/8-th wavelength implies that both correlations in~\eqref{eq:corrs1} are $1/\sqrt{2}$.} to both $\h$ and~$\partial_x\eta$ (i.e.~onto the $\cos$ and $\sin$). Indeed,
\begin{subequations}
\begin{align}
	\corr{(\hop,\bnabla\bcdot\bE)} \approx \corr{(\partial_x\hop, \bnabla\bcdot\bE)} \approx 0.69\com\label{eq:corrs1}
\end{align}
where $\corr(a,b)\defn\<ab\>\big/\sqrt{\<a^2\>\<b^2\>}$. For topography $\hcl$ in figure~\ref{fig:divE}(b) these correlations are at least an order of magnitude less:
\begin{align}
	\corr{(\hcl,\bnabla\bcdot\bE)} \approx 0.01 \ \text{and} \ \corr{(\partial_x\hcl,\bnabla\bcdot\bE)} =-0.07\per\label{eq:corrs2}
\end{align}
\end{subequations}



\section{The stability of the lower- and upper-branch solutions for topography $\hop$\label{sec:stability}}

For topography of the form $\h=\h_0\cos{(mx)}$ (as is~\eqref{def_hop}) we can understand both the transitions among the three flow regimes, as well as the role the transient eddies play on eddy saturation by studying a 3-dimensional invariant manifold of model~\eqref{EOM}.

It is easy to verify in this case that there exists a 3-dimensional invariant manifold by assuming that the eddy streamfunction is:
\begin{align}
  \psi =\[S(t)\,\sin{(mx)}+C(t)\,\cos{(mx)}\]/m\per\label{psi3state}
\end{align}
Inserting~\eqref{psi3state} into~\eqref{EOM} the Jacobian term~$\J(\psi,\h)$ vanishes identically, and thus~\eqref{EOM} reduce to:
\begin{subequations}
	\begin{align}
		\frac{\dd C}{\dd t} &= -\mu C + m(\beta/m^2- U) S \com\\
		\frac{\dd S}{\dd t} &= -\mu S - m(\beta/m^2- U) C - \eta_0 U\com\\
		\frac{\dd U}{\dd t} &= F -\mu U + \half\eta_0  S\com
	\end{align}\label{EOM3state}\end{subequations}
(after also ignoring $\nu_4$ in $\D$). Steady equilibrium solutions of~\eqref{EOM3state} can be found in the form:
\beq
	U=U^e\ ,\ \ \psi =\underbrace{\[S^e\,\sin{(mx)}+C^e\,\cos{(mx)}\]/m}_{\defn\psi^e}\per\label{eq:UeCeSe}
\eeq
Both the lower- and upper-branch solutions for the case with topography $\hop$ are \emph{exactly} such steady solutions in the form of~\eqref{eq:UeCeSe}; see panels (b) and~(d) of figure~\ref{spec_open}.

{\color{black}
Starting with the work by \citet{Hart-1979}, the stability of solutions~\eqref{eq:UeCeSe} to perturbations that lie within this invariant manifold has been extensively studied \citep{Hart-1979,Pedlosky-1981,Kallen-1982,Rambaldi-Flierl-1983,Yoden-1985}.  Here, we study the stability of~\eqref{eq:UeCeSe} to any general perturbation that may or may not lie within the invariant manifold. \citet{Charney-Flierl-1980}  studied the stability of the inviscid and unforced version of~\eqref{eq:eig_phi}, i.e., with $\mu=F=0$ and assuming the large-scale mean flow~$U^e$ to be a given external parameter of the problem. Here, we perform the stability of the forced--dissipative problem. Thus, we use~$U^e$ that results from a steady solutions of~\eqref{EOM} (or, for $\hop$, a steady solution of~\eqref{EOM3state}). We investigate how this instability may elucidate why eddy saturation is observed.}

%

\subsection{Stability calculation}
Assume perturbations about the steady state~\eqref{eq:UeCeSe}:
\ifdraft
\begin{align}
	U(t) &= U^e+\upsilon(t) \ ,\ \
	\psi(x,y,t) &= \psi^e(x,y) + \phi(x,y,t) + [ s(t)\,\sin{(mx)}+ c(t)\,\cos{(mx)}]/m \per\label{pertUeCeSe}
\end{align}
\else
\beq
\begin{aligned}
	U(t) &= U^e+\upsilon(t)\com\\
	\psi(x,y,t) &= \psi^e(x,y) + \phi(x,y,t)\\
	&\hspace{2em} + [ s(t)\,\sin{(mx)}+ c(t)\,\cos{(mx)}]/m \per
\end{aligned}\label{pertUeCeSe}
\eeq
\fi
In~\eqref{pertUeCeSe} we write the perturbation streamfunction as a sum of a perturbation that lies within the 3-state invariant manifold $[s\,\sin{(mx)}+ c\,\cos{(mx)}]/m$ and a perturbation $\phi$ that does not project on the invariant manifold. That is, $\phi$ satisfies: $\langle \phi\,\ee^{\ii m x}\rangle=0$.

By inserting~\eqref{pertUeCeSe} into~\eqref{EOM} we get the linearized equations for the perturbations. When projected on the invariant manifold the perturbation equations read:\begin{subequations}	\begin{align}
	\frac{\dd c}{\dd t} &= -\mu c + m(\beta/m^2- U^e) s- m S^e\,\upsilon \com\\
	\frac{\dd s}{\dd t} &= -\mu s - m(\beta/m^2- U^e) c - (\eta_0 - m  C^e)\upsilon\com\\
	\frac{\dd \upsilon}{\dd t} &= -\mu \upsilon + \half\eta_0\,s\com
\end{align}\label{eq:stab_3d}\end{subequations}
while the orthogonal complement of the perturbation equations is:
\ifdraft
\begin{align}
&\p_t\lap\phi = - \[(U^e\lap+\beta)\p_x +\mu\lap\] \phi -\left\{ \vphantom{\lap_{n_y}} [S^e\cos{(m x)}-C^e\sin{(m x)}](\lap+m^2)+ m\h_0\sin{(m x)}\vphantom{\lap_{n_y}}\right\} \p_y\phi \per \label{eq:eig_phi}
\end{align}\else
\begin{align}
 &\p_t\lap\phi = - \[(U^e\lap+\beta)\p_x +\mu\lap\] \phi -\left\{ \vphantom{\lap_{n_y}} [S^e\cos{(m x)}\right.\nonumber\\&\quad\left.-C^e\sin{(m x)}](\lap+m^2)+ m\h_0\sin{(m x)}\vphantom{\lap_{n_y}}\right\} \p_y\phi \per
\label{eq:eig_phi}
\end{align}\fi
Eigenanalysis of~\eqref{eq:stab_3d} determines the stability of steady state~\eqref{eq:UeCeSe} to perturbations within the invariant manifold; eigenanalysis of~\eqref{eq:eig_phi} determines the stability of~\eqref{eq:UeCeSe} to perturbations outside the manifold. {\color{black}Details for the eigenanalysis of~\eqref{eq:eig_phi} are provided in the~Appendix.}


\begin{figure*}
\centerline{\includegraphics[width=33pc]{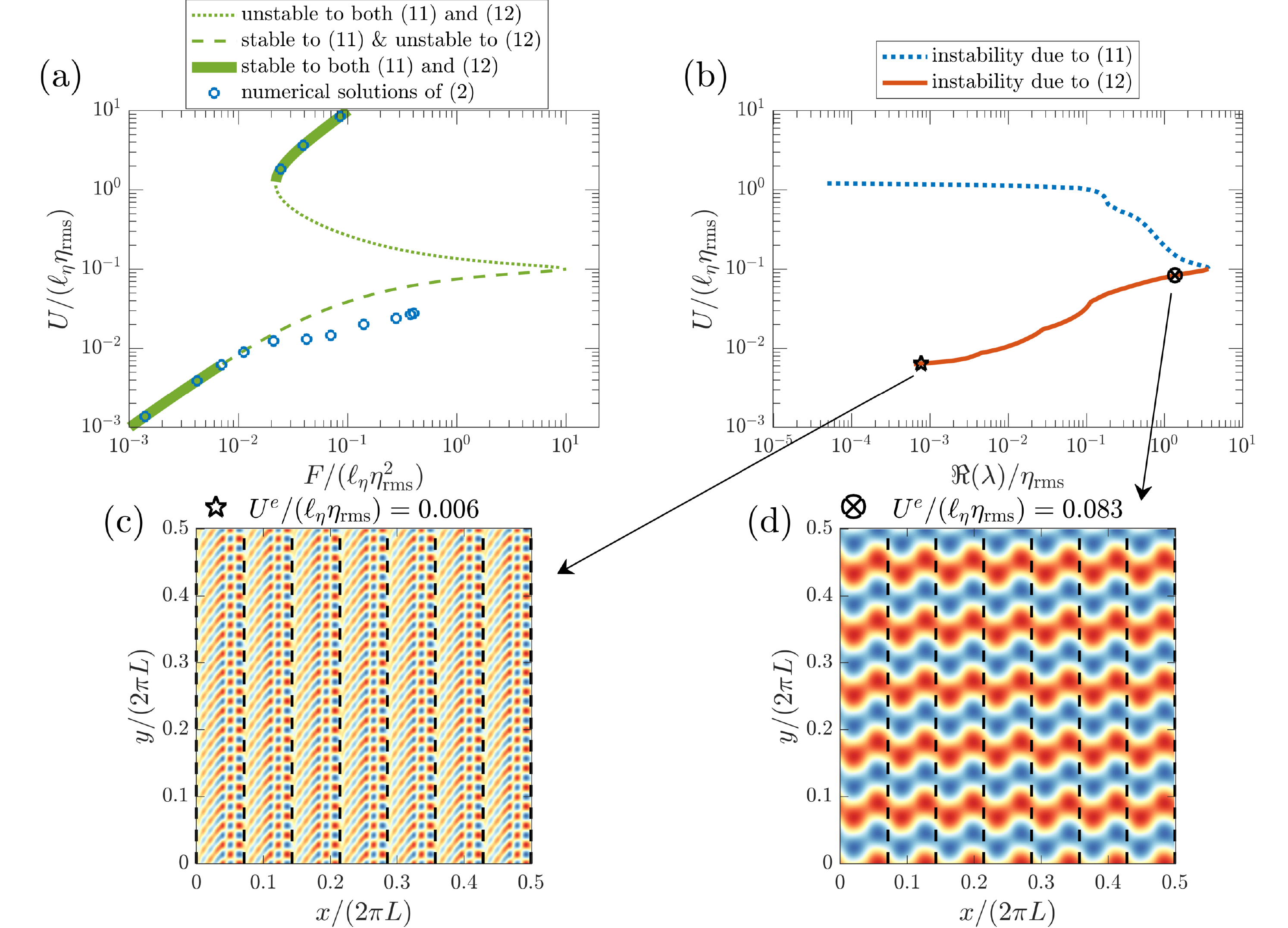}}
\caption{Panel (a) shows the steady solutions~\eqref{eq:UeCeSe} for topography $\hop$ and their stability as predicted by~\eqref{eq:stab_3d} and~\eqref{eq:eig_phi}. Also shown for comparison are the numerical solutions of~\eqref{EOM} with circles. Panel (b) shows the instability growth rates, i.e, the real part of the eigenvalue $\lambda$, according to~\eqref{eq:stab_3d} and~\eqref{eq:eig_phi}. Only unstable eigenvalues are shown. The structure of the most unstable eigenfunction that correspond to $U^e$ just above the marginal point for instability is shown in panel~(c), while the eigenfunction for $U^e$ just before maximum instability is shown in panel~(d) (these are the two cases marked in panel~(b)). The eigenfunction in panel~(c) has $n_y=2.5m$ and Bloch wavenumber $n_x=m/2$; the eigenfunction in panel~(d) has $n_y=m$ and Bloch wavenumber $n_x=0$. The dashed black lines mark the topographic PV wavelength. Notice that only part of the domain is show for clarity. (The eigenfunctions were obtained with eigenanalysis of~\eqref{eq:eig_Mny}.)}\label{fig:stability}
\end{figure*}

\begin{figure}
 \centerline{\includegraphics[width=17pc]{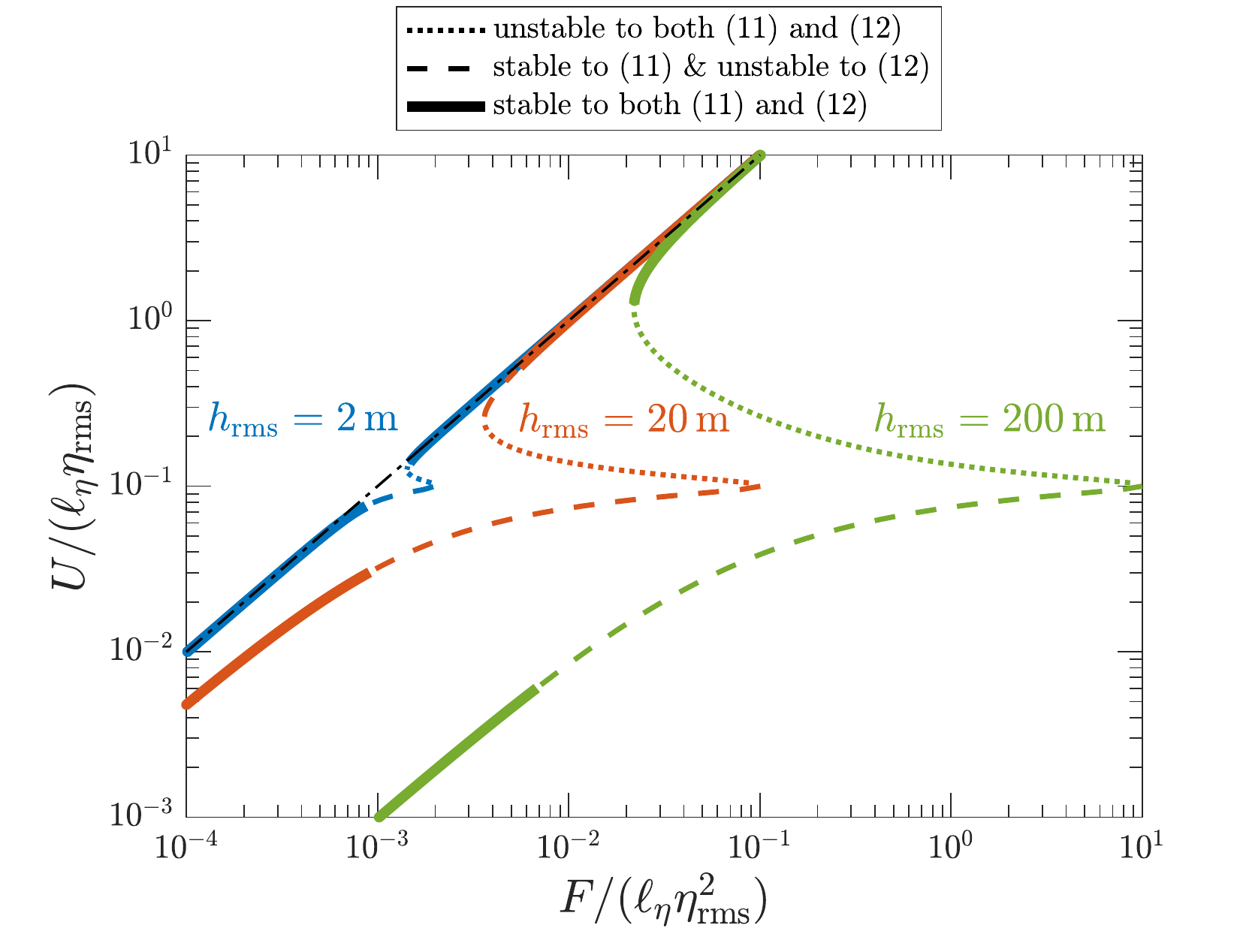}}
 \caption{Steady solutions~\eqref{eq:UeCeSe} for topography $\hop$ with varying r.m.s.~topographic height. Also shown is their stability as predicted by~\eqref{eq:stab_3d} and~\eqref{eq:eig_phi}. The range of $F$ values for which there is instability scales roughly as~$h_{\rm rms}^2$. Dash-dotted curve marks the flat-bottom solution $F/\mu$.}\label{fig:stability_hrms}
\end{figure}

\subsection{Stability results}

{\color{black}
We can understand the various flow transitions for topography $\hop$ by studying the stability of steady states~\eqref{eq:UeCeSe}.  The region of stability of~\eqref{eq:UeCeSe}, both within and outside the low-dimensional manifold (i.e. with respect to both~\eqref{eq:stab_3d} and~\eqref{eq:eig_phi}), is marked with the thick semitransparent curve in figure~\ref{U_F_comb}(a). The onset of this barotropic--topographic instability explains: \emph{(i)}~the appearance of transient eddies as wind stress is increased and \emph{(ii)}~the termination of the upper-branch solution as wind stress is decreased.

}


Figure~\ref{fig:stability}(a) shows the large-scale flow $U^e$ of the steady states~\eqref{eq:UeCeSe} as a function of wind stress forcing $F$, together with their stability as predicted by~\eqref{eq:stab_3d} and~\eqref{eq:eig_phi}. Also shown in figure~\ref{fig:stability}(a) are the numerical results of~\eqref{EOM} for comparison. It is clear that the transition from the lower branch to the turbulent regime, in which eddy saturation is observed, is triggered by the instability due to~\eqref{eq:eig_phi} rather than by~\eqref{eq:stab_3d}. This instability of the lower branch first occurs at $\Fs=7.15\times10^{-3}$. Figure~\ref{fig:stability}(c) shows the most unstable eigenfunction just beyond the marginal point of instability. This eigenfunction has small-scale meridional structure; the meridional wavenumber of the eigenfunction shown in figure~\ref{fig:stability}(c) is $n_y=2.5m$. Thus, the instability of the lower branch due to~\eqref{eq:stab_3d} introduces $y$-dependence in the flow; see figure~\ref{fig:stability}(c).

Next, we would like to investigate how this instability contributes to the occurrence of eddy saturation. {\color{black}What is interesting here is how the instability maximum growth} rate varies with the large-scale flow $U^e$. Figure~\ref{fig:stability}(b) shows that in the regime which we find eddy saturation the instability maximal growth rate due to~\eqref{eq:stab_3d} increases dramatically with $U$; growth rates increase by a factor of $10^3$ when $U$ increases only by a factor of~15. The transient eddy source, thus, increases significantly for only small changes in $U$. In this way, and by analogy with arguments for baroclinic eddy saturation, larger wind stress forcing implies the need for stronger eddies and this leads to eddy saturation. Furthermore, this suggests that the flow adjusts to a state close to marginal stability for this barotropic--topographic instability, similarly to the baroclinic marginal stability argument first invoked by~\citet{Stone-1978}.

\ifdraft\else\vspace{1em}\fi
The stability calculations presented in this section are based on that the topography does not depend on $y$. However, the same flow transitions occur for flows with open geostrophic contours above complex topographies. In that case, finding the lower- and upper-branch equilibrium steady states is much more painful since some of the Jacobian terms, for example $\J(\psi^e,\h)$, are \emph{not} identically zero. Therefore, the lower- and upper-branch solutions do not lie within any low-dimensional manifold. Methods for obtaining such equilibria were developed by \citet{Tung-Rosenthal-1985}.

\section{Discussion}

The results reported here demonstrate that eddy saturation can occur even without baroclinicity. A bare barotropic setting is capable of producing {\color{black}an eddy saturation regime in which the transport is insensitive to wind stress forcing. The model shows additionally some other features of eddy saturation that were previously observed in more elaborate ocean models: EKE varies close to linearly with wind stress forcing (instead of quadratically) and also transport increases with increasing bottom drag.}

{\color{black} In this barotropic model with no baroclinic eddies, the flow relies on the barotropic--topographic instability to produce transient eddies.} For the $y$-independent topography with open geostrophic contours $\hop$, we extended the stability analysis of~\citet{Hart-1979} to encompass any general flow perturbations. Thus, we managed to identify the various flow transitions shown in figure~\ref{U_F_comb}(a) with the onset of an instability. The stability calculations presented section~\ref{sec:stability} crucially depend on that the topography $\hop$ does not depend on~$y$. However, in the context of this barotropic model, we expect that the same flow transitions above more complex topography $\h(x,y)$ with open geostrophic contours would result from a similar barotropic--topographic instability. {\color{black} More importantly, by studying the barotropic--topographic instability we showed that for the case with eddy saturation the transient eddy source (i.e.~the barotropic--topographic instability) had a particular property. The growth rate of the instability increases dramatically with small changes of the large-scale flow. This explains why eddy saturation occurs: an increase of the wind stress strength requires the generation of more transient eddies to carry this excess of momentum to the solid Earth through form stress; but only a slight change in the transport can give rise to the transient eddies needed to balance the momentum. This is analogous with arguments for baroclinic eddy saturation.

The factor which controls the appearance or not of eddy saturated states in a barotropic setting is whether or not the geostrophic contours are open. This was demonstrated here using two simple sinusoidal topographies that bare this distinction. However, the bare existence of the eddy saturation in this simple setting depends on the presence of a transient-eddy source (i.e. of an instability). The range of wind stress forcing values that eddy saturation could potentially occur is limited by the range of wind stress forcing values for which the barotropic--topographic instability occurs. And the latter crucially depends on the height of the topography. For the flat-bottom case with $h=0$ all geostrophic contours are open but there is no eddy saturation since there is no instability. Figure~\ref{fig:stability_hrms} shows the steady states~\eqref{eq:UeCeSe} together with their stability (similarly to figure~\ref{fig:stability}(a)) for various topography heights. The range of wind stress forcing values that we have instability scales roughly as $h_{\rm rms}^2$.
}


We note, that the flow characteristics that were described in section~\ref{sec:results} are not quirks of the simple sinusoidal topographies~\eqref{topos}. The eddy saturation regime, the drag crisis, and multiple equilibria that are present here when geostrophic contours are open, have been all also recently found to exist in this model for a flow above a random monoscale topography with open geostrophic contours \citep{Constantinou-Young-2017} and also above a multi-scale topography that has a $k^{-2}$ power spectrum (not reported).

In conclusion, the results presented here emphasized that barotropic processes might play a role in shaping the zonal momentum balance in the ACC and in producing eddy saturation. In particular, barotropic processes are responsible at least for the component of the transport that is related to the bottom flow velocity of the ACC. (This bottom-velocity component of the ACC transport accounts for about 25\% of the total transport according to \citet{Donohue-etal-2016}.) Even though the ACC is strongly affected by baroclinic processes, our results here reinforce the increasing evidence arguing for the importance of the barotropic mechanisms in determining the ACC transport \citep{Ward-Hogg-2011,Thompson-NaveiraGarabato-2014,Youngs-etal-2017}. {\color{black}Additionally, the results of this paper argue that topography does not only have a passive role in the momentum budget acting merely as a sink for zonal momentum, but it can also have an active role by producing transient eddies that shape the standing eddy field and its form stress.}

\acknowledgments
I am mostly grateful to William Young for his support and his insightful comments. Discussions with Anand Gnanadesikan, Petros Ioannou, Spencer Jones, Sean Haney, Cesar Rocha, Andrew Thompson, Gregory Wagner, and Till Wagner are greatly acknowledged. Also, I thank Louis-Philippe Nadeau and David Straub for their constructive review comments. This project was supported by the NOAA Climate and Global Change Postdoctoral Fellowship Program, administered by UCAR's Cooperative Programs for the Advancement of Earth System Sciences.

\appendix

\appendixtitle{Stability of steady states~\eqref{eq:UeCeSe} to perturbations outside the low-dimensional manifold}\label{app:stab_phi}

For the eigenanalysis of~\eqref{eq:eig_phi}, first note that if $\phi$ is independent of $y$ then~\eqref{eq:eig_phi} implies stability. Therefore, a necessary condition instability of~\eqref{eq:eig_phi} is $\p_y\phi\ne0$. In this case, also note that:
\begin{enumerate}
	\item[\emph{(i)}]\label{remark1}
	\eqref{eq:eig_phi} is homogeneous in $y$, i.e., if $\phi(x,y,t)$ is a solution so is $\phi(x,y+a,t)$ for any $a$, and
	\item[\emph{(ii)}]\label{remark2}
	the coefficients of~\eqref{eq:eig_phi} remain unchanged under $x\mapsto x+ 2\pi \kappa/m$ for any integer $\kappa$, i.e.,~if $\phi(x,y,t)$ is a solution so is $\phi(x+ 2\pi \kappa/m,y,t)$.
\end{enumerate}
Remark~\emph{(i)} implies that the linear operator in~\eqref{eq:eig_phi} commutes with the translation operator in $y$ and therefore they share the same eigenfunctions. Thus, the eigensolution has $y$-dependence $\propto \ee^{\ii n_y y}$. Remark~\emph{(ii)} implies that the linear operator in~\eqref{eq:eig_phi} commutes with the appropriate translation operator by $2\pi \kappa/m$ in $x$, and thus the eigensolution has an $x$-dependence in the form of a Bloch wavefunction.\footnote{For a more detailed discussion on Bloch eigenfunctions the reader is referred, for example, to the textbook by~\citet{Ziman-1979} or to the Appendix~D in the thesis of~\citet{Constantinou-2015-phd}.} Therefore, we search for the eigensolutions of~\eqref{eq:eig_phi} as:
\beq
	\phi(x,y,t) = \ee^{\lambda t} \ee^{\ii n_y y}\tilde{\phi}_{n_x}(x)\com\label{eq:phiansantz}
\eeq
with
\begin{align}
\tilde{\phi}_{n_x} \defn  \ee^{\ii n_x x}\sum_{M=-\infty}^{+\infty} c_M \ee^{\ii  M m x}\per\label{eq:tildephi}
\end{align}
In~\eqref{eq:phiansantz} it is understood that the eigenvalue $\lambda$ depends on both~$n_x$ and~$n_y$. In~\eqref{eq:tildephi}, $n_x$ is the zonal Bloch wavenumber and it is restricted to take values $|n_x|\le m/2$ (in condensed matter physics this is referred to as the first Brillouin zone). To see why $n_x$ is restricted, consider what~\eqref{eq:tildephi} (or equivalently remark~\emph{(ii)}) implies for the zonal spectral power of~$\phi$. Eigenfunction $\phi$ cannot have spectral power in any arbitrary zonal wavenumber, but instead it can only have power at zonal wavenumbers:
\beq
\pm n_x,\, \pm(n_x\pm m),\, \pm(n_x\pm 2m),\, \dots\per\label{eq:blochwaves}
\eeq
From~\eqref{eq:blochwaves} we can verify that Bloch wavenumbers $n_x$ and $\pm(n_x-m)$ are completely equivalent and thus we only need to study the stability for $|n_x|\le m/2$ (cf.~\citet{Ziman-1979,Constantinou-2015-phd}).

With the ansantz~\eqref{eq:phiansantz} the stability of~\eqref{eq:eig_phi} reduces to a generalized eigenvalue problem for each meridional wavenumber $n_y$:
\begin{align}
\lambda  \lap_{n_y} \tilde{\phi}_{n_x} = \Mop_{n_y} \tilde{\phi}_{n_x}\com\label{eq:eig_Mny}
\end{align}
where $\lap_{n_y} \defn \partial^2_x-n_y^2$ and
\ifdraft
\begin{align}
	\Mop_{n_y}&\defn-[(U^e\lap_{n_y} +\beta)\partial_x + \mu\lap_{n_y}]- \ii n_y \left\{ [S^e\cos{(m x)}-C^e\sin{(m x)} ](\lap_{n_y}+m^2) + m\h_0\sin{(m x)}\vphantom{\lap_{n_y}}\right\}\per\label{eig:Lop}
\end{align}
\else
\begin{align}
	\Mop_{n_y}&\defn-[(U^e\lap_{n_y} +\beta)\partial_x + \mu\lap_{n_y}]- \ii n_y \left\{ [S^e\cos{(m x)}\vphantom{\lap_{n_y}}\right. \nonumber\\
	&\quad \left. -C^e\sin{(m x)} ](\lap_{n_y}+m^2) + m\h_0\sin{(m x)}\vphantom{\lap_{n_y}}\right\}\per\label{eig:Lop}
\end{align}
\fi
The problem is further reduced if we insert~\eqref{eq:tildephi} into~\eqref{eq:eig_Mny}. Then, the eigenproblem~\eqref{eq:eig_Mny} reduces into an infinite system of equations whose unknowns are the coefficients $c_M$ and the eigenvalue~$\lambda$. Truncating the sum in~\eqref{eq:tildephi} up to $|M|\le M_{\max}$ leaves us with $2 M_{\max}+1$ equations for $2 M_{\max}+2$ unknowns: $c_{-M_{\max}},c_{-M_{\max}+1},\dots,c_{M_{\max}}$ and $\lambda$. These equations are compactly written as:
\ifdraft
\beq
\Aop\(\lambda\)\[c_{-M_{\max}},\dots,c_{M_{\max}}\]^{\transp} = 0\com\label{Ac=0}
\eeq
\else
\beq
\Aop\(\lambda\) \begin{bmatrix}
         c_{-M_{\max}} \\ \vdots \\ c_{M_{\max}}
        \end{bmatrix} = 0\com\label{Ac=0}
\eeq
\fi
where $\Aop$ is a $(2 M_{\max}+1)\times(2 M_{\max}+1)$ matrix. A non-trivial solution of~\eqref{Ac=0} exists only when $\det[\Aop\(\lambda\)]=0$, and this last condition determines the eigenvalues $\lambda$ that correspond to the eigenfunction with wavenumbers~$n_x$ and~$n_y$ \citep{Lorenz-1972,Gill-1974,Charney-Flierl-1980}.



\end{document}